\documentclass[11pt]{article}%
\usepackage{srcltx}
\usepackage{amsfonts}
\usepackage{amssymb,amsbsy}
\usepackage{graphicx}%

\newtheorem{theo}{Theorem}
\newtheorem{cor}{Corollary}
\newtheorem{prop}{Proposition}
\newtheorem{defin}{Definition}
\newcommand{\be}{\begin{equation}}
\newcommand{\ee}{\end{equation}}
\newcommand{\ba}{\begin{array}}
\newcommand{\ea}{\end{array}}
\newcommand{\p}{\partial}

\newcommand{\ds}{\displaystyle}

\date{}
\begin{document}

\title{\protect\vspace{-1.5cm}{\Large
Coupling constant metamorphosis as an
integrability-preserving transformation for general finite-dimensional dynamical systems and ODEs}
\protect\vspace{-0.2cm}
}

\author{{\sc Artur Sergyeyev}\\[3mm]
Mathematical Institute, Silesian University in Opava,\\ Na
Rybn\'\i{}\v{c}ku 1, 746\,01 Opava, Czech Republic\\ E-mail: {\tt
Artur.Sergyeyev@math.slu.cz}} \maketitle

\begin{abstract}\protect\vspace{-1cm}
In the present paper we extend the multiparameter coupling constant metamorphosis, also known as the generalized St\"ackel transform, from Hamiltonian dynamical systems to general finite-dimensional
dynamical systems and ODEs. This transform
interchanges the values of integrals of motion with the parameters
these integrals depend on but leaves the phase space
coordinates intact. Sufficient conditions under which
the transformation in question preserves integrability and a simple
formula relating the
solutions of the original system to those of the transformed one
are given.\looseness=-1

\medskip

\end{abstract}

\protect\vspace{-0.5cm}
\section{Introduction}
The modern theory of integrable finite-dimensional dynamical
systems concentrates mostly %(albeit not exclusively)
on Hamiltonian systems. For the latter, the presence of symplectic or
Poisson structure combined with the Liouville theorem leads to a
number of important simplifications, in particular regarding the
relationship among symmetries and integrals of motion. However,
this also leads to certain restrictions because while considering
various transformations of Hamiltonian systems one naturally wants
the transformed systems to be Hamiltonian too.
%There certainly is
%no need to recall that
Such transformations are of interest for two reasons: they enable
one to reduce the lists of integrable systems of certain form
resulting from various classification procedures, and, sometimes,
to relate new integrable systems to the known ones.\looseness=-1

In the present paper we explore a somewhat surprising situation
when the {\em absence} of need to preserve the Poisson or symplectic
structure considerably {\em simplifies} things for general
(non-Hamiltonian) dynamical systems including ODEs.
%, instead of making them more complicated.
This is precisely the case for the multiparameter coupling constant metamorphosis,
also known as the generalized St\"ackel transform, see \cite{hiet,
km, sb08, bls2011} and references therein. The said transform maps
%gives a recipe for transforming
a set of integrals of motion in involution
into another such set. In essence, it interchanges the {\em parameters}
present in the integrals of
motion with the {\em values} of those integrals, thus producing a new set of
integrals of motion. Note that the dynamical variables, i.e., the
phase space coordinates, are {\em not} affected by this
transformation.\looseness=-1

The above recipe works provided the original set of integrals of
motion depends on some parameters in a nontrivial fashion, and
enables one to produce a number of interesting new examples or,
conversely, to relate certain new integrable systems to the known
ones, cf.\ \cite{sb08, bls2011}. However, in the Hamiltonian case
there is a catch: the associated
transformation for the equations of motion can only be written
down upon restricting these equations to a common level surface of
integrals of motion generating the transform under study
and turns out to be a rather nontrivial
reciprocal transformation, see \cite{sb08} for details. We need
this reciprocal transformation precisely because we want the
transformed equations of motion to originate from the transformed
Hamiltonian through the Poisson structure which should be left
intact.\looseness=-1

Below we extend the multiparameter generalized St\"ackel transform
to general continuous dynamical systems, for which there is no
Poisson structure to preserve. In this case it is possible to
jettison the reciprocal transformation described above and define
the associated transformation of equations of motion in a much
simpler fashion.
%The same applies to solutions and symmetries of the system in question.
These results are summarized in Theorem~\ref{mgstt} and
Corollary~\ref{mgstc} for general continuous dynamical systems and
the overdetermined systems of first order PDEs, respectively, and
in Corollary~\ref{mgstt4odes} for ODEs. What is more,
Proposition~\ref{lax} shows that the transformed system inherits
existence of a Lax representation from the original one.
The most important advantage of abandoning the reciprocal
transformations %as outlined above
is a very simple relationship among solutions of the original
system and those of the transformed one, see Remark~1 below.
\looseness=-1

As an aside, note that the requirement of presence of parameters
in the dynamical systems under study and in their integrals of motion
is not as restrictive as it may seem at first glance because the
parameters can often be introduced by hand, e.g.\ through changes
of dependent and independent variables, and often simple
transformations like translation or rescaling of dependent
variables enable one to produce interesting examples, see e.g.
Examples~2, 3 and 5 below.\looseness=-1

\section{Coupling constant metamorphosis
for general dynamical systems}\label{sec2}

Consider an open domain $M\subset \mathbb{K}^n$
($\mathbb{K}=\mathbb{R}$ or $\mathbb{C}$) and a
%(multiparametric nonstationary)
dynamical system on $M$,
\begin{equation}\label{ourds}
d x^\alpha/dt =X^\alpha(t,x^1,\dots,x^n,a_1,\dots,a_k),\quad
\alpha=1,\dots,n,
\end{equation}
where $x^\alpha$ are coordinates on $M$, and $a_j\in\mathbb{K}$
are parameters. Thus, {\em de facto} we have a $k$-parametric
family of dynamical systems but for the ease of writing we shall
refer to it below as if it were a single dynamical system. The
same convention will apply to its symmetries, integrals of motion,
Lax matrices, etc. We deliberately use the local setting instead
of the more global one (vector fields on manifolds, etc.) because
of the subsequent necessity to invoke the implicit function theorem which
almost inevitably forces one to consider things locally, as
explained below. In what follows all objects are tacitly assumed
to be smooth in all of their arguments.\looseness=-1

To (\ref{ourds}) we can naturally associate a vector field on $M$
which depends on the parameters $t$, $a_1,\dots,a_k$,
\[
X=\sum\limits_{\alpha=1}^n
X^\alpha\displaystyle\frac{\partial}{\partial x^\alpha}.
\]

Recall that a (smooth) function
$f=f(t,x^1,\dots,x^n,a_1,\dots,a_k)$ is an {\em integral of
motion} (or a first integral) for the dynamical system
(\ref{ourds}) if we have
\begin{equation}\label{intdef}
\p f/\p t + X(f)=0.
\end{equation}

Consider another dynamical system on $M$,
\begin{equation}\label{dssym}
d x^\alpha/d\tau =Y^\alpha(t,x^1,\dots,x^n,a_1,\dots,a_k),\quad
\alpha=1,\dots,n,
\end{equation}
and the associated vector field on $M$,
\[
Y=\sum\limits_{\alpha=1}^n
Y^\alpha\displaystyle\frac{\partial}{\partial x^\alpha}.
\]
Recall that $Y$ is a {\em symmetry} for (\ref{ourds}) if we have
\[
\displaystyle\frac{d^2 x^\alpha}{dt d\tau}=\frac{d^2
x^\alpha}{d\tau dt}, \quad\alpha=1,\dots,n,
\]
where the derivatives are computed by virtue of (\ref{ourds}) and
(\ref{dssym}), or equivalently,
\begin{equation}\label{symdef}
\p Y/\p t+[X,Y]=0.
\end{equation}
Here and below $[\cdot,\cdot]$ stands for the Lie bracket of
vector fields (i.e., the usual commutator of differential
operators) unless otherwise explicitly stated.

%Likewise, a vector field $Y$ is a {\em symmetry} for $X$ (or for the dynamical system associated with $X$) %if its Lie derivative along $X$ vanishes: $\mathcal{L}_X (Y)=[X,Y]=0$, where $[\cdot,\cdot]$ stands for the %Lie bracket of vector fields. \looseness=-1

%Now assume that our vector field $X$ depends on $k$ parameters $a_1,\dots,a_k$. Then, in general, its %integrals depend on these parameters too.

Let (\ref{ourds}) have $k$ functionally independent integrals
$I_1,\dots,I_k$ such that
\begin{equation}
\label{det}\det\left(|\!|\p I_{i}/\p
a_j|\!|_{i,j=1,\dots,k}\right)\neq 0.
\end{equation}

Then by the implicit function theorem the equations
\be\label{idi}
I_i(x,t, a_1,\dots,a_k)=b_i,\quad i=1,\dots,k, \quad x\in M,
\ee
where
$b_i$ are constants, are (in general only locally, see the discussion
in the beginning of the next section) uniquely solvable w.r.t.
$a_i$. Denote the solution in question as follows:
\be\label{sol}
a_i=\tilde I_i(x,t,b_1,\dots,b_k),\quad i=1,\dots,k,\quad x\in M.
\ee
The reason for this notation will become clear in a moment.

%Let $\tilde X$ be a vector field obtained from $X$ by substituting $\tilde I_i$ for $a_i$ for all
%$i=1,\dots,k$. More broadly,
If $K$ is a geometrical object on $M$ (a function, a vector field,
a tensor field, a differential form, etc.) which may depend on the
parameters $a_i$, then $\tilde K$ will stand for the geometrical
object obtained from $K$ by substituting $\tilde I_i$ for $a_i$
for all $i=1,\dots,k$. We shall write this as
\[
\tilde K=\left.K\right|_{a_1=\tilde I_1, \dots, a_k=\tilde I_k}.
\]

Thus, for instance, $\tilde X$ is a vector field obtained from $X$
by substituting $\tilde I_i$ for $a_i$ for all $i=1,\dots,k$:
\[
\tilde X=\sum\limits_{\alpha=1}^n
\left.X^\alpha\right|_{a_1=\tilde I_1, \dots, a_k=\tilde
I_k}\displaystyle\frac{\partial}{\partial x^\alpha},
\]
and the associated dynamical system reads
\begin{equation}\label{trds}
d x^\alpha/dt =\tilde
X^\alpha(t,x^1,\dots,x^n,b_1,\dots,b_k),\quad \alpha=1,\dots,n.
\end{equation}

The only exception from the above notational convention is made
for the functions $\tilde I_i$ which are not related to $I_i$ in the
fashion described above. The reason for this apparent discrepancy
is that $\tilde I_i$ turn out to be first integrals for $\tilde X$, see
Theorem~\ref{mgstt} below.

%Indeed, it is immediate that $\tilde I_i$ can be considered as functions of $I_j$, $j=1,\dots,k$ and of $b_l$, %$l=1,\dots,k$. As $b_l$ are constants, to prove that $\tilde X(\tilde I_i)=0$, $i=1,\dots,k$ it suffices to show that %$\tilde X(I_i)=0$, $i=1,\dots,k$, but the latter equalities readily follow from our assumption that
%$X(I_i)=0$, $i=1,\dots,k$.
%Indeed, we have

Note that $\tilde X$ and $\tilde I_i$ depend on the parameters
$b_1,\dots,b_k$, and that
$\tilde I_i$ obviously are functionally independent. %as  the implicit function theorem.

In analogy with \cite{sb08}, we shall refer to the procedure of
passing from (\ref{ourds}), $X$ and $I_i$ to (\ref{trds}), $\tilde
X$ and $\tilde I_i$ as to the $k$-{\em parameter generalized
St\"ackel transform} (or as to the $k$-{\em parameter coupling constant metamorphosis})
generated by $I_1,\dots,I_k$. We shall also
say that (\ref{ourds}) is {\em St\"ackel-equivalent} to
(\ref{trds}).\looseness=-1

Just as in the Hamiltonian setting \cite{sb08}, we
have the following duality: when applied to (\ref{trds}), the
$k$-parameter generalized St\"ackel transform generated by
$\tilde{I}_1,\dots, \tilde{I}_k$ brings us back to the integrals
$I_1,\dots, I_k$ and the system (\ref{ourds}) we have started
with. \looseness=-1

As a final remark, note that the above construction admits
%lends itself, at least locally, to
the following geometric interpretation.

Consider the extended phase space $N=M \times P$, where $P\subset \mathbb{K}^k$,
an open subset of $\mathbb{K}^k$, is
the space where our parameters live: $P\ni \vec a=(a_1,\dots,a_k)^T$;
here and below the superscript $T$ indicates the transposed matrix.
%Obviously, $N$ is a trivial foliation of codimension $k$,
%which we denote $\mathcal{F}_{k}$, of the form $N=\bigcup\limits_{\vec a\in P} M\times
%\lbrace\vec a\rbrace$, where $
%\lbrace\vec a\rbrace$ denotes a one-element set consisting of a single vector $\vec a$.
\looseness=-1

The dynamical system (\ref{ourds}) can be naturally extended to $N$ upon
setting
\begin{equation}\label{ourdsext}
d a_i/d t=0,\quad i=1,\dots,k.
\end{equation}
In other words, the extended dynamical system (\ref{ourds})+(\ref{ourdsext}) on $N$ is determined by
{\em the same} vector field $X$ as the original system (\ref{ourds}) but
$X$ is now treated as a vector field on $N$. Conversely,
the original system (\ref{ourds})
is recovered from the extended one upon {\em fixing} the values
of parameters $a_i$, $i=1,\dots,k$.

The extended system under study admits, in addition to $k$ integrals of motion
$I_j$, $j=1,\dots,k$, which are there by assumption,
the `obvious' integrals of motion $a_j$, $j=1,\dots,k$ (for a moment, we ignore
the possibility of existence of further integrals of motion for (\ref{ourds})).
Thus, $N$ foliates, at least locally, into common level surfaces
of these $2k$ integrals. We
shall denote this foliation by $\mathcal{F}_{2k}$.\looseness=-1

Roughly speaking, %for our extended dynamical system
the $k$-parameter generalized
St\"ackel transform, defined above,
interchanges $a_j$ and $b_j$ as well as $I_j$ and $\tilde I_j$,
%in other words, we change the functional basis in the
%algebra of integrals of motion for our extended dynamical system
%generated by $a_j$ and $I_j$ from the original one, i.e., $a_j, I_j$, $j=1,\dots,k$,
%to $b_j, \tilde I_j$, $j=1,\dots,k$,
i.e., we choose a different way to parameterize the
leaves of $\mathcal{F}_{2k}$.

The transform under study turns the {\em extended} system (\ref{ourds})+(\ref{ourdsext}) into (\ref{trds})+(\ref{trdsext}),
where (\ref{trdsext}) reads
\begin{equation}\label{trdsext}
d b_i/d t=0,\quad i=1,\dots,k.
\end{equation}
While (\ref{ourds})+(\ref{ourdsext}) and (\ref{trds})+(\ref{trdsext}) coincide, at least locally (cf.\ the assumptions regarding the applicability of the implicit function theorem in the next section), on any given leaf of $\mathcal{F}_{2k}$, they are {\em different}
when considered on $N$ as a whole.\looseness=-1

%Geometrically, we introduce a new (and possibly
%defined only locally, i.e., we may have to assume
%$N$ to be sufficiently small for things to work out well)
%foliation $\mathcal{F}'_k$ of $N$
%of codimension $k$ instead of $\mathcal{F}_k$:
%$N=\bigcup\limits_{\vec b\in \tilde P} \widetilde{M}\times
%\lbrace\vec b\rbrace$, where $\vec b=(b_1,\dots,b_k)^T$, $\widetilde{M}$ is a suitable %open domain in $\mathbb{K}^n$, and $\tilde P$ is an open domain in $\mathbb{K}^k$
%where %the new vector parameter
%$\vec b$ takes values. Recall that $b_j$ are related to $a_j$ via (\ref{idi}), and
%thus the foliation $\mathcal{F}'_k$ is different from $\mathcal{F}_k$.
\looseness=-1

%However, if we fix a leaf $\mathcal{L}=\{a_i=a_i^0, I_i=b_i^0,i=1,\dots,k\}$ of %$\mathcal{F}_{2k}$ (here $a_i^0, b_i^0\in\mathbb{K}$, $i=1,\dots,k$), then the transform %under study, roughly speaking, just reparameterizes this leaf $\mathcal{L}=\{\tilde %I_i=a_i^0, b_i=b_i^0,i=1,\dots,k\}$, and the {\em restriction}
%of (\ref{ourds})+(\ref{ourdsext}) onto this leaf remains essentially unchanged;
%this becomes especially obvious if we use $a_j$ and $I_j$,
%$j=1,\dots,k$, as a part of the local coordinate system on $N$.

Our primary interest is, however, in the dynamical systems on the
{\em original} phase space $M$ rather than on the whole $N$,
% or on a leaf $\mathcal{L}$,
and this is where things
become even more nontrivial:
%while on $N$
%a fixed leaf $\mathcal{L}$ of $\mathcal{F}_{2k}$
%we basically just switch the
%roles of parameters and integrals, as explained above,
the transformed dynamical system (\ref{trds})
on $M$ arises upon fixing the {\em new} parameters $b_j$
rather than the old ones $a_j$,
%i.e., upon picking a leaf of the foliation
%$\mathcal{F}'_k$ rather than that of $\mathcal{F}_k$,
and
(\ref{trds}) is a restriction onto $M$ of the {\em transformed} extended dynamical system (\ref{trds})+(\ref{trdsext}) on $N$.\looseness=-1
%In other words, (\ref{trds}) arises as a {\em new} restriction on $M$
%of the extended dynamical system associated with the original vector
%field $X$.

\section{Main results}

The above considerations suggest
that the generalized St\"ackel transform should preserve
a number of integrability attributes of the original system (e.g.\ symmetries,
integrals of motion, etc.), and in the present section we state and prove
the relevant results.\looseness=-1

Here and below we use the following {\em blanket assumption}. We
suppose that the domain $M$ and the ranges of values of time $t$
and of the parameters $a_j$ and $b_j$, $j=1,\dots,k$, are chosen
so that the implicit function theorem ensures that the system
(\ref{idi}) has a unique solution with respect to $a_i$,
$i=1,\dots,k$. In general this means making $M$ and the ranges in
question sufficiently small because of the local nature of the
implicit function theorem, but we deliberately do not fully spell
out here the relevant conditions as there exists a number of
situations when they would be too restrictive, e.g.\ when $I_j$ are
linear in all $a_j$, $j=1,\dots, k$ (or, more broadly, when we
have explicit formulas for $\tilde I_j$ and one can employ these
to specify $M$ and the relevant ranges, as it is the case for the
majority of interesting examples).\looseness=-1

\begin{theo}\label{mgstt}
%Under the above assumptions the following
%holds
%For all $x\in W$, all $t\in\tilde L$ and all $(b_1,\dots,b_k)\in \tilde B$ we have:
Let the dynamical system (\ref{ourds}) have $k$ functionally
independent integrals $I_i$ such that (\ref{det}) holds. Define
$\tilde I_i$ and the transformed quantities like $\tilde X$ as
above.

Then the following assertions hold:

i) the functions $\tilde I_i$, $i=1,\dots,k$, are functionally
independent integrals for (\ref{trds}), and we have
\begin{equation}
\label{det2}\det\left(|\!|\p \tilde I_{i}/\p
b_j|\!|_{i,j=1,\dots,k}\right)\neq 0;
\end{equation}

ii) if $J_1,\dots,J_m$ is another set of integrals for
(\ref{ourds}) such that all integrals $I_1,\dots,I_k$, $J_1,\dots,
J_m$ are functionally independent, then
%, and let $\tilde J_s$ be obtained
%from $J_s$ by substituting $\tilde I_i$ for $a_i$ for all $i=1,\dots,k$.
%Then
$\tilde I_1,\dots,\tilde I_k, \tilde J_1,\dots, \tilde J_m$ are
functionally independent integrals for (\ref{trds});

iii) if $Y_1,\dots, Y_r$ are linearly independent symmetries for
(\ref{ourds}) such that \be\label{symint}Y_p (I_j)=0\quad\mbox{for
all}\quad p=1,\dots,r\quad\mbox{and}\quad j=1,\dots,k, \ee then
$\tilde Y_1,\dots, \tilde Y_r$ are linearly independent symmetries
for (\ref{trds}), and \be\label{symint1}\tilde Y_p (\tilde
I_j)=0\quad\mbox{for all}\quad p=1,\dots,r\quad\mbox{and}\quad
j=1,\dots,k; \ee
%%%$\tilde Y_i (\tilde I_j)=0$ for all $i=1,\dots,r$.

iv) if under the assumptions of (ii) and (iii) the symmetries $Y_1, \dots,
Y_s$, where $s\leq r$, span an involutive distribution, i.e., $[Y_p,Y_q]=\sum\limits_{g=1}^s
c_{pq}^{g}(a_1,\dots,a_k, I_1,\dots,I_k,J_1,\dots,J_m) Y_g$ for all $p,q=1,\dots,s$,
%and for some
%$c_{pq}^s\in\mathbb{K}$ (the latter are assumed to be independent of $a_1,\dots,a_k$),
then the symmetries $\tilde Y_1, \dots,
\tilde Y_s$ also span an involutive distribution, $[\tilde Y_p, \tilde
Y_q]=\sum\limits_{g=1}^s \tilde c_{pq}^{g}\tilde Y_g$, for all
$p,q=1,\dots,s$, where $\tilde c_{pq}^{g}=c_{pq}^{g}(\tilde I_1,\dots,\tilde I_k,\allowbreak b_1,\dots,b_k,\tilde
J_1,\dots,\tilde J_m)$. If $c_{pq}^g\in\mathbb{K}$ are constants (in particular, they do not depend on $a_j$,
$j=1,\dots,k$), and thus
$Y_g$, $g=1,\dots,s$, form a Lie algebra, then $\tilde Y_g$, $g=1,\dots,s$, form an isomorphic Lie
algebra.\looseness=-1 \end{theo}

{\em Proof.}
Consider the identities \be\label{idi1} b_i\equiv I_i(x,\tilde
I_1,\dots,\tilde I_k),\quad i=1,\dots,k, \quad x\in M, \ee that
follow from (\ref{sol}).
Obviously, $\partial b_i/\partial t+\tilde X(b_i)\equiv 0$, as
$b_i$ are constants, and hence acting by $\partial/\partial
t+\tilde X$ on the left-hand side and the right-hand side of
(\ref{idi1}) yields
\[
\ba{lcl}
%\begin{aligned}
0&\equiv &\left.\left(\displaystyle\frac{\partial
I_i}{\partial t}+\tilde X(I_i)\right)\right|_{a_1=\tilde I_1,
\dots, a_k=\tilde I_k} +\sum\limits_{j=1}^k
\left(\displaystyle\frac{\partial \tilde I_j}{\partial t}+\tilde X(\tilde
I_j)\right)\left.\left(\ds\frac{\p I_i}{\p a_j}\right)\right|_{a_1=\tilde
I_1, \dots, a_k=\tilde I_k}\\[5mm]
%\]
%\[
%\ba{l}
&=&\ds\left.\left(\displaystyle\frac{\partial I_i}{\partial
t}
+X(I_i)\right)\right|_{a_1=\tilde I_1, \dots, a_k=\tilde I_k}
+\sum\limits_{j=1}^k
\left(\displaystyle\frac{\partial \tilde I_j}{\partial t}+\tilde X(\tilde
I_j)\right)\left.\left(\ds\frac{\p I_i}{\p a_j}\right)\right|_{a_1=\tilde
I_1, \dots, a_k=\tilde I_k}\\[5mm]
&=&\ds\sum\limits_{j=1}^k
\left(\displaystyle\frac{\partial \tilde I_j}{\partial t}
+\tilde X(\tilde
I_j)\right)\left.\left(\ds\frac{\p I_i}{\p a_j}\right)\right|_{a_1=\tilde
I_1, \dots, a_k=\tilde I_k}, \quad i=1,\dots,k.
\ea
%\end{aligned}
\]
Here we used the chain rule and the fact that $I_i$ are integrals
for (\ref{ourds}), and therefore by (\ref{intdef}) we have
$$\displaystyle\frac{\partial I_i}{\partial t}+X(I_i)=0.$$

Using (\ref{det}) we readily find that $\partial\tilde
I_j/\partial t +\tilde X(\tilde I_j)=0$, $j=1,\dots, k$, so
$\tilde I_j$ are indeed integrals for (\ref{trds}).\looseness=-1

Eq.(\ref{det2}) follows from the implicit function theorem. The
functional independence of $\tilde I_j$ is immediate. This
completes the proof of part (i).

Next, $I_j$ and $J_s$
are integrals of motion for (\ref{ourds}) by assumption, so $\partial
I_j/\partial t+X(I_j)=0$ and $\partial J_s/\partial t+ X(J_s)=0$,
and using the chain rule shows that $\tilde J_s$
are integrals of motion for (\ref{trds}) as we have
\[
\ba{l}
\ds\frac{\partial \tilde J_s}{\partial t}+\tilde X(\tilde J_s)=\left.\left(\ds\frac{\partial J_s}{\partial t}+X(J_s)
+ \sum\limits_{j=1}^k \left(\frac{\partial I_j}{\partial t}+
X(I_j)\right)\ds\frac{\p J_s}{\p a_j}\right)\right|_{a_1=\tilde I_1,
\dots, a_k=\tilde I_k}=0,\quad s=1, \dots, m. \ea
\]
The functional independence of $\tilde I_1,\dots,\tilde I_k$,
$\tilde J_1,\dots, \tilde J_m$ easily follows from that of
$I_1,\dots, I_k$, $J_1,\dots, J_m$, and thus part (ii) is also
proven.

In a similar fashion, further taking into account (\ref{symint})
and bearing in mind that $\partial Y_q/\partial t+[X,Y_q]=0$ by
assumption as $Y_q$ are symmetries for (\ref{ourds}), we obtain
\be\label{symcomm0} \ba{l}
\hspace*{-5mm}\displaystyle\frac{\partial\tilde Y_q}{\partial
t}+[\tilde X,\tilde
Y_q]\!=\!\!\!\left.\left(\!\displaystyle\frac{\partial
Y_q}{\partial
t}+[X,Y_q]+\!\!\sum\limits_{j=1}^k\!\left(\!\!\left(\!\displaystyle\frac{\partial
I_j}{\partial t}+X(I_j)\right)\frac{\p Y_q}{\p
a_j}-Y_q(I_j)\frac{\p X}{\p
a_j}\!\right)\!\!\!\right)\!\right|_{a_1=\tilde I_1, \dots,
a_k=\tilde I_k}\hspace{-17mm}=0,\quad q=1, \dots, r, \ea \ee as
desired. The linear independence of $\tilde Y_q$ readily follows
from that of $Y_q$. Let us stress that if we drop the condition
(\ref{symint}), the right-hand side of (\ref{symcomm0}) is in
general no longer obliged to vanish, and hence the quantities
$\tilde Y_i$ will no longer be symmetries for $\tilde X$, cf.\
Example 1 below.

Finally, to prove (iv) we note that, in complete analogy with
(\ref{symcomm0}), we have
\[
\begin{array}{l}
[\tilde Y_p, \tilde
Y_q]=\left.\left([Y_p,Y_q]+\sum\limits_{j=1}^k\left(\displaystyle
Y_p(I_j)\frac{\p Y_q}{\p a_j}-Y_q(I_j)\frac{\p Y_p}{\p
a_j}\right)\right)\right|_{a_1=\tilde I_1, \dots, a_k=\tilde
I_k}\\[5mm]
=\ds\left.[Y_p,Y_q]\right|_{a_1=\tilde I_1, \dots, a_k=\tilde
I_k} =\sum\limits_{s=1}^r \left.\left(c_{pq}^{s}Y_s\right) \right|_{a_1=\tilde
I_1, \dots, a_k=\tilde I_k} =\sum\limits_{s=1}^r \tilde c_{pq}^{s}\tilde
Y_s,\quad p,q=1, \dots, r,
\end{array}
\]
and the result follows. $\square$

Informally, Theorem~\ref{mgstt} states that if $I_i$ and $J_s$ are
integrals and $Y_j$ are symmetries for (\ref{ourds}), and
(\ref{det}) and (\ref{symint}) hold, then $\tilde I_i$ and $\tilde
J_s$ are integrals and $\tilde Y_j$ are symmetries for
(\ref{trds}), the Lie algebra of symmetries $\tilde Y_j$ is
`essentially isomorphic' to that of $Y_j$, and (\ref{det2}) and (\ref{symint1})
hold. By a slight abuse of terminology, it can be said that the
multiparameter coupling constant metamorphosis (or the generalized St\"ackel transform) preserves integrals
of motion and symmetries that respect the generators $I_j$ of
the transform in question. Proceeding in the spirit of the proof
of Theorem~\ref{mgstt} it can be shown that the
multiparameter generalized St\"ackel transform preserves invariant
curves and surfaces, Darboux multipliers, Jacobi multipliers and
other similar structures. Thus, under certain technical assumptions
the transformed system (\ref{trds}) inherits the integrability
properties of (\ref{ourds}).

It is now appropriate to
recall the definition of
extended integrability due to Bogoyavlenskij \cite{bo}:\looseness=-1
\begin{defin}[\cite{bo}]\label{dbi}
A dynamical system (\ref{ourds}) is {\em integrable in the broad
sense} if it has $m$ functionally independent integrals of motion
$J_1,\dots,J_m$, where $n>m\geq 0$, and $n-m$ linearly independent
commuting symmetries $Y_1,\dots,Y_{n-m}$ such that $Y_i(J_j)=0$
for all $i=1,\dots,n-m$ and all $j=1,\dots,m$.
\end{defin}

Theorem~\ref{mgstt} implies that the generalized St\"ackel
transform preserves extended integrability. Namely, the following
assertion holds.

\begin{cor}\label{extint}
Let (\ref{ourds}) be integrable in the broad sense, with the
integrals $J_j$ and symmetries $Y_p$ as in Definition~\ref{dbi}.
Consider a $k$-parameter generalized St\"ackel
transform generated by the integrals $I_j$ which are functions of
$J_s$, $s=1,\dots,m$, i.e., $I_j=I_j(J_1,\dots,J_m)$, $j=1,\dots,k$,
and assume that (\ref{det}) holds.

Then the transformed system (\ref{trds}) is again integrable in
the broad sense.

\end{cor}

{\em Proof.} To prove this corollary it suffices to notice that we
can construct from $J_1,\dots,J_m$ a new set of functionally
independent integrals, say, $I_s$, $s=1,\dots,m$, for
(\ref{ourds}), so that $I_j$ for $s\leq k$ are precisely the generators of the
St\"ackel transform in question. Then by
Theorem~\ref{mgstt} the transformed quantities $\tilde I_s$,
$s=1,\dots,m$, and $\tilde Y_j$, $j=1,\dots,n-m$, meet the
requirements of Definition~\ref{dbi}
%the above definition of extended integrability
for (\ref{trds}), if so do $I_s$ and $Y_j$ for (\ref{ourds}), and the
result follows. $\square$ \looseness=-1

{\bf Remark 1.} Unlike the case of Hamiltonian dynamical systems
considered in \cite{sb08}, where the reciprocal transformation was
involved, we have a very simple recipe for relating the solutions
of (\ref{ourds}) to those of (\ref{trds}). Namely, if
$x^\alpha=\Xi^\alpha(t,a_1,\dots,a_k)$, $\alpha=1,\dots,n$, is a
solution for (\ref{ourds}) then \be\label{soltr}
x^\alpha=\left.\Xi^\alpha(t,a_1,\dots,a_k)\right|_{t=\tilde t,
a_1=\tilde I_1, \dots, a_k=\tilde I_k}, \alpha=1,\dots,n, \ee
is an {\em implicit}
solution for (\ref{trds}).

In particular, if the formulas
$$x^\alpha=\Xi^\alpha(t,a_1,\dots,a_k,C_1,\dots,C_n),\quad
\alpha=1,\dots,n, $$ where $C_1,\dots,C_n$ are arbitrary
constants, define a general solution for (\ref{ourds}) then
the formulas \be\label{soltr1}
\left.x^\alpha=\Xi^\alpha(t,a_1,\dots,a_k,C_1,\dots,C_n)\right|_{t=\tilde
t, a_1=\tilde I_1, \dots, a_k=\tilde I_k},\quad \alpha=1,\dots,n,
\ee define an implicit general solution for (\ref{trds}). Using
similar considerations one can also readily find out how an implicit or
parametric (general or particular) solution of (\ref{ourds}) transforms into
an implicit or parametric (general or particular) solution of (\ref{trds}).

%It is important to
As a final remark, note that the multiparameter
generalized St\"ackel transform preserves (the existence of) the
Lax representations. Namely, it is readily verified that the
following assertion holds.

\begin{prop}\label{lax}
Let (\ref{ourds}) admit a Lax representation of the form
$dL/dt=[M,L]$, where $L$ and $M$ are $N\times N$ matrices that
depend on $t,x^1,\dots,x^n,a_1,\dots,a_k$ and on a spectral
parameter $\lambda$, and $[,]$ stands here for the commutator of
matrices.

Then the transformed dynamical system (\ref{trds}) possesses a Lax
representation of the form $d\tilde{L}/dt=[\tilde M,\tilde L]$,
where
\[
\ba{l} \tilde
L=\left.L(t,x^1,\dots,x^n,a_1,\dots,a_k,\lambda)\right|_{a_1=\tilde
I_1, \dots, a_k=\tilde I_k},\quad % \\[2mm]
\tilde M=\left.M(t,x^1,\dots,x^n,a_1,\dots,a_k,\lambda)\right|_{a_1=\tilde
I_1, \dots, a_k=\tilde I_k}. \ea
\]
\end{prop}

\section{Examples}
To illustrate the above results, we start with the following easy example.

{\bf Example 1.} Consider the one-component
nonstationary dynamical system (i.e., a first-order ODE),
\be\label{ex1ds} dx/dt=ax/((x-t)^2+a x), \ee see equation
1.4.3-2.15 in \cite{zp}.

For now let us work over $\mathbb{C}$. Then (\ref{ex1ds}) admits
an integral $I=\ln x+a/(t-x)$,
and hence is integrable in the broad sense.

Setting $a_1\equiv a$ and $b_1\equiv b$ we readily find that
$\tilde I=(b-\ln x)(t-x)$
is an integral of motion for the transformed equation
\[
d x/d t=(b-\ln x)/(x-t+b-\ln x),
\]
which therefore is also integrable in the broad sense.

As we have already noticed in Introduction, while many dynamical
systems of interest do not involve parameters, we can often
introduce the parameters `by hand', e.g., through translation or
rescaling of dependent variables.

{\bf Example 2.} Consider the following dynamical system from
Example 2.22 of \cite{gor}
\[
dx/dt=-2x^2 + 2z,\quad dy/dt=-3xy,\quad dz/dt=4xz- 2x(2x^2 -9y^2).
\]
which has an integral of motion of the form $z-x^2+3y^2$.

Upon rescaling the variable $y$, $y\rightarrow (a/3)^{1/2} y$, we
obtain the system
\be\label{dse1} dx/dt=-2x^2 + 2z,\quad
dy/dt=-3xy,\quad dz/dt=4xz- 2x(2x^2 -3 a y^2)
\ee
with an integral
of motion
\[
I=z- x^2+ a y^2.
\]
Setting $a_1\equiv a$ and $b_1\equiv b$ and applying the general
theory presented above we find that
\[
\tilde I=\frac{b+x^2-z}{y^2}
\]
is an integral of motion for the transformed system
\be\label{trs1} dx/dt=-2x^2 + 2z,\quad dy/dt=-3xy,\quad
dz/dt=-2 xz +2x^3 +6 b x. \ee
We now see that in the transformed system the
right-hand side of the third equation is independent of $y$, so we
have a decoupled subsystem for $x$ and $z$,
\[
dx/dt=-2x^2 + 2z,\quad dz/dt=-2 xz +2x^3 +6 b x,
\]
i.e., (\ref{trs1}) is, in a sense, indeed a somewhat simpler
object than the original system (\ref{dse1}). \looseness=-1
Moreover, (\ref{trs1}) admits a symmetry $Y=y\p/\p y$. However, as
$Y$ does not preserve $\tilde I$, $Y(\tilde I)=-2\tilde I \neq
0$, this symmetry has no counterpart for the original system
(\ref{dse1}).

On the other hand, (\ref{trs1}) is easily seen to have
another integral $\tilde I_2=x^2/2+z
+2b \ln(y)$, and hence (\ref{trs1}) is integrable by quadratures
(and integrable in the broad sense).

Indeed, upon restriction onto
the common level surface $\tilde I=C_1$ and $\tilde I_2=C_2$
the system (\ref{trs1}) boils down to a single ODE,
\[
dy/dt=\mp y\sqrt {6\left(C_1-b+C_2 y^{2}-2 b\ln y\right)},
\]
which is obviously integrable by quadratures,
and hence so is (\ref{trs1}).

Now, by Corollary~\ref{extint} the above implies that (\ref{dse1}) is
also integrable in broad sense (and integrable by quadratures,
as we can readily obtain the general solution for (\ref{dse1})
from that of (\ref{trs1}) using (\ref{soltr}) and (\ref{soltr1})).
Note that the counterpart of $\tilde I_2$ for (\ref{dse1}) reads
\[
I_2=x^2/2+z
+2(a y^2-x^2+z) \ln(y)
\]

{\bf Example 3.} For a somewhat more elaborated example, consider
system 9.25 from \cite{kamke}
\[
du/dt=- u v^2+u+v,\quad dv/dt=u^2 v -u-v,\quad dw/dt=v^2-u^2
\]
which  has two integrals of motion, $u^2+v^2+\ln w^2$ and $w(u
v-1)$.

Upon rescaling $u\rightarrow (a_1)^{1/2} u$, $v\rightarrow
(a_1)^{1/2} v$, $w\rightarrow w/a_2$, we obtain the system
\[
du/dt=- a_1 u v^2+u+ v,\quad dv/dt= a_1 u^2 v -u-v,\quad dw/dt=a_1
a_2 (v^2-u^2)
\]
with the integrals of motion
\[
I_1=a_1 (u^2+w^2) +\ln (w^2),\quad I_2=a_2 w(a_1 u v-1).
\]

Consider the two-parametric generalized St\"ackel transform
generated by $I_1$ and $I_2$. We find that
\[
\tilde I_1 = (b_1-\ln(w^2))/(u^2+w^2),\quad \tilde I_2 = -b_2
(u^2+w^2)/(u v w(\ln(w^2)-b_1)+(u^2+w^2) w)
\]
are integrals of motion for the system
\[
\ba{rcl} du/dt&=&- (b_1-\ln(w^2))u v^2/(u^2+w^2) +u+ v,\\[2mm]
dv/dt&=& (b_1-\ln(w^2))u^2 v/(u^2+w^2)  -u-v,\\[2mm] dw/dt&=&b_2
(b_1-\ln(w^2))(u^2-v^2)/(u v w(\ln(w^2)-b_1)+(u^2+w^2) w). \ea
\]

\section{Generalized St\"ackel transform for overdetermined
partial differential systems}
%Another important observation

%As a final remark,
Theorem~\ref{mgstt} admits a natural
generalization to the overdetermined systems of first-order PDEs
which naturally arise e.g.\ in the study of zero-curvature
representations, pseudopotentials and B\"acklund transformations
for integrable (systems of) PDEs, cf.\ e.g.\ \cite{as} and references therein.

Namely, consider an overdetermined system of first-order PDEs of
the form
%(Pfaff system)
\begin{equation}\label{dspdes}
\displaystyle\frac{\partial x^\alpha}{\partial t^A}
=X_A^\alpha(t^1,\dots,t^d,x^1,\dots,x^n,a_1,\dots,a_k),\quad
\alpha=1,\dots,n,\quad A=1,\dots,d,
\end{equation}
and assume that this system is in involution, i.e., \be\label{inv}
\displaystyle\frac{\partial^2 x^\alpha}{\partial t^A \partial
t^B}=\frac{\partial^2 x^\alpha}{\partial t^B \partial t^A},
\quad\alpha=1,\dots,n,\quad A,B=1,\dots,d, \ee where the
derivatives are computed by virtue of (\ref{dspdes}), or
equivalently,
\begin{equation}\label{symdef1}
\partial X_A/\partial t^B-\partial X_B/\partial t^A-[X_A,X_B]=0, \quad A,B=1,\dots,d.
\end{equation}

\begin{cor}\label{mgstc}
Let (\ref{inv}) hold, and let (\ref{dspdes}) have $k$ {\em joint}
(i.e., such that $\partial I_j/\partial t^A+X_A(I_j)=0$ for all
$A$ and $j$) functionally independent integrals $I_1,\dots,I_k$
such that (\ref{det}) is satisfied.

Then the following assertions hold:

i) the vector fields $\tilde X_A$, $A=1,\dots,d$, again commute:
$[\tilde X_B, \tilde X_B]=0$, $a,b=1,\dots,d$, and hence the
transformed system
\[
\displaystyle\frac{\partial x^\alpha}{\partial t^A} =\tilde
X_A^\alpha(t^1,\dots,t^d,x^1,\dots,x^n,b_1,\dots,b_k),\quad
\alpha=1,\dots,n,\quad A=1,\dots,d,
\]
is again in involution;

ii) the functions $\tilde I_i$, $i=1,\dots,k$, are functionally
independent joint integrals for the vector fields $\tilde X_A$,
$a=1,\dots,d$, and we have
%\begin{equation}
%\label{det2}
\[
\det\left(|\!|\p \tilde I_{i}/\p
b_j|\!|_{i,j=1,\dots,k}\right)\neq 0;
\]
%\end{equation}

iii) if $J_1,\dots,J_m$ is another set of joint integrals for
$X_A$, $A=1,\dots,d$, such that all integrals
$I_1,\dots,I_k,\allowbreak J_1,\dots,\allowbreak J_m$ are
functionally independent, then
$\tilde I_1,\dots,\tilde I_k,\allowbreak \tilde
J_1,\dots,\allowbreak \tilde J_m$ are joint functionally
independent integrals for $\tilde X_A$, $A=1,\dots,d$;

iv) if $Y_1,\dots, Y_r$ are linearly independent %over $C^\infty(M)$
joint (i.e., $\partial Y_q/\partial t^A+[X_A, Y_q]=0$ for all $A$
and $q$) symmetries for $X_A$, $A=1,\dots,d$, such that
%\be\label{symint}
\[
Y_p (I_j)=0\quad\mbox{for all}\quad
p=1,\dots,r\quad\mbox{and}\quad j=1,\dots,k,
%\ee
\]
then $\tilde Y_1,\dots, \tilde Y_r$ are linearly independent joint
symmetries for $\tilde X_A$, $A=1,\dots,d$, and
%\be\label{symint1}
\[
\tilde Y_p (\tilde I_j)=0\quad\mbox{for all}\quad
p=1,\dots,r\quad\mbox{and}\quad j=1,\dots,k;
\]
%\ee
%%%$\tilde Y_i (\tilde I_j)=0$ for all $i=1,\dots,r$.

v) if under the assumptions of (iii) and (iv) the symmetries $Y_1, \dots,
Y_s$, where $s\leq r$, span an involutive distribution, i.e., $[Y_p,Y_q]=\sum\limits_{g=1}^s
c_{pq}^{g}(a_1,\dots,a_k, I_1,\dots,I_k,J_1,\dots,J_m) Y_g$ for all $p,q=1,\dots,s$,
then the symmetries $\tilde Y_1, \dots,
\tilde Y_s$ also span an involutive distribution, i.e., $[\tilde Y_p, \tilde
Y_q]=\sum\limits_{g=1}^s \tilde c_{pq}^{g}\tilde Y_g$, for all
$p,q=1,\dots,s$, where $\tilde c_{pq}^{g}=c_{pq}^{g}(\tilde I_1,\dots,\tilde I_k, \allowbreak b_1,\dots,b_k,\tilde
J_1,\dots,\tilde J_m)$. If $c_{pq}^g\in\mathbb{K}$ are constants (in particular, they do not depend on $a_j$,
$j=1,\dots,k$), and thus
$Y_g$, $g=1,\dots,s$, form a Lie algebra, then $\tilde Y_g$, $g=1,\dots,s$, form an isomorphic Lie
algebra.\looseness=-1 \end{cor}
Stating the counterpart of Proposition~\ref{lax} for
(\ref{dspdes}) is left as an exercise for the reader.

\section{Applications to ODEs}

Consider an ODE resolved with respect to the highest-order
derivative: \be\label{ode0} d^m u/d z^m=F\left(z, u, du/dz,\dots,
d^{m-1} u/d z^{m-1},a_1,\dots,a_k\right). \ee Let $n=m$, and put
\be\label{ode2ds} x^1=u,\quad x^2=du/dz,\quad \dots,\quad
x^{m}=d^{m-1} u/d z^{m-1}. \ee Consider a dynamical system
\be\label{dsode0} \ba{l}
d x^1/d t=x^2,\quad d x^2/d t=x^3, %\quad  d x^3/d t=x^4,
\quad\dots, d x^{m-1}/d t=x^{m},\quad  d x^{m}/d t=f(t,
x^1, \dots, x^{m},a_1,\dots,a_k);
\ea \ee %where
\[
\mbox{here}\ f(t, x^1, \dots, x^{m},a_1,\dots,a_k) =\left.
F\left(z, u,\ds\frac{du}{dz},\dots, \frac{d^{m-1} u}{d
z^{m-1}},a_1,\dots,a_k\right)\right|_{z=t, u=x^1, du/dz=x^2,
\dots, d^{m-1} u/d z^{m-1}=x^{m}}.
\]
It is well known that the dynamical system (\ref{dsode0}) is
equivalent to (\ref{ode0}), and we can readily apply the result of
Theorem~\ref{mgstt} to (\ref{dsode0}).

What is more, it is immediate that upon applying the
multiparameter generalized St\"ackel transform to (\ref{dsode0})
we obtain the system of the same kind, that is, \be\label{dsode1}
\ba{l}
d x^1/d t=x^2,\quad d x^2/d t=x^3, %\quad  d x^3/d t=x^4,
\quad\dots,\\[3mm] d x^{m-1}/d t=x^{m},\quad d x^{m}/d t=\tilde
f(t, x^1, \dots, x^{m},b_1,\dots,b_k), \ea \ee which is, through
(\ref{ode2ds}), equivalent to an ODE of the form
\be\label{ode1}
\ds\frac{d^m u}{d z^m}=\tilde F\left(z, u,
\ds\frac{du}{dz}, \dots, \frac{d^{m-1} u}{d
z^{m-1}},b_1,\dots,b_k\right), \ee
where
\be\label{ds2ode} \ba{l} \tilde
F\left(\ds z, u, \frac{du}{dz},\dots, \frac{d^{m-1} u}{d
z^{m-1}},b_1,\dots,b_k\right)=\left.\tilde f(t, x^1,
\dots, x^{m+1}, b_1, \dots, b_k)\right|_{t=z, x^1=u, x^2=\frac{du}{dz},
\dots, x^{m}=\frac{d^{m-1} u}{d z^{m-1}}}. \ea \ee

Thus, %at the end of the day
we have obtained a transformation relating the ODEs (\ref{ode0})
and (\ref{ode1}), and this transformation preserves the
integrability properties.

In view of the particular interest in the study of ODEs let us
restate Theorem~\ref{mgstt} for this special case directly in
terms of ODEs. To this end we first recall the relevant
definitions following \cite{olv}.

A generalized vector field $Y=h(z,u,du/dz,\dots,d^{m-1}u/d
z^{m-1})\p/\p u$ is a {\em (generalized) symmetry} for
(\ref{ode0}) if we have
\[
D^m (h)-\sum\limits_{j=0}^{m-1} \ds\frac{\p F}{\p u_j} D^j (h)=0.
\]
Here $u_0\equiv u, u_j\equiv d^j u/d z^j$, and we have introduced
the so-called operator of the total $z$-derivative
\[
D=\ds\frac{\p}{\p z}+F\frac{\p}{\p
u_{m-1}}+\sum\limits_{j=0}^{m-2}u_{j+1}\frac{\p}{\p u_j}
\]
(here we
treat $z$ and $u_j$ as formally independent entities, see e.g.\
\cite{olv} for details).

Also, a function
$I=I(z,u,du/dz,\dots,\allowbreak d^{m-1}u/d z^{m-1})$ is a {\em
(first) integral} for (\ref{ode0}) if $D(f)=0$.

It is easily seen that upon passing from (\ref{ode0}) from
(\ref{dsode0}) an integral of motion $I$ and the {\em
prolongation} (see e.g.\ \cite{olv}) of a symmetry $Y$
\[
\mathrm{pr}\, Y=\sum\limits_{j=0}^{m-1} D^j(h) \frac{\p}{\p u_j}
\]
become respectively an integral and a symmetry for (\ref{ode0}) in
the sense of the definitions from Section~\ref{sec2}.\looseness=-1

If $Y_i=h_i(z,u,du/dz,\dots,d^{m-1}u/d z^{m-1})\p/\p u$, $i=1,2$,
%and $Y_2=h_2(z,u,du/dz,\dots, \allowbreak  d^{m-1}u/d z^{m-1})\p/\p u$
are two symmetries for (\ref{ode0}) in the sense of the above
definition, their commutator is \cite{olv} given by the formula
\begin{equation}\label{com_ode}
[Y_1,Y_2]=\left(\mathrm{pr}\, Y_1 (h_2)-\mathrm{pr}\, Y_2
(h_1)\right)\p/\p u,.
\end{equation}
and of course it is again a symmetry for (\ref{ode0}).

With all this in mind we are ready to state the ODE version of
Theorem~\ref{mgstt}.

\begin{cor}\label{mgstt4odes}
Under the above assumptions, let (\ref{ode0}) be an ODE whose
right-hand side depends on $k$ parameters $a_1,\dots,a_k$, and let
(\ref{ode0}) have $k$ functionally independent integrals
$I_1,\dots,I_k$ such that (\ref{det}) holds.\looseness=-1

Then the following claims hold:
%for all $(u,du/dz,\dots, d^{m-1}u/d z)\in W$, all $z\in\tilde L$
%and all $(b_1,\dots,b_k)\in \tilde B$:

i) the functions $\tilde I_i$, $i=1,\dots,k$, are functionally
independent integrals for the transformed ODE (\ref{ode1}), and we
have
%\begin{equation}\label{det2_ode}
\[
\det\left(|\!|\p \tilde I_{i}/\p
b_j|\!|_{i,j=1,\dots,k}\right)\neq 0;
\]
%\end{equation}

ii) if $J_1,\dots,J_m$ is another set of integrals for
(\ref{ode0}) such that all integrals $I_1,\dots,I_k$, $J_1,\dots,
J_m$ are functionally independent, then
%, and let $\tilde J_s$ be obtained
%from $J_s$ by substituting $\tilde I_i$ for $a_i$ for all $i=1,\dots,k$.
%Then
$\tilde I_1,\dots,\tilde I_k, \tilde J_1,\dots, \tilde J_m$ are
functionally independent integrals for (\ref{ode1});

iii) if $Y_1,\dots, Y_r$ are linearly independent generalized
symmetries for (\ref{ode0}) such that
%\be\label{symint_ode}
\[
\mathrm{pr}\, Y_p (I_j)=0\quad\mbox{for all}\quad
p=1,\dots,r\quad\mbox{and}\quad j=1,\dots,k,
\]
%\ee
then $\tilde Y_1,\dots, \tilde Y_r$ are linearly independent
generalized symmetries for (\ref{ode1}), and
%\be\label{symint1_ode}
\[
\mathrm{pr}\, \tilde Y_p (\tilde I_j)=0\quad\mbox{for all}\quad
p=1,\dots,r\quad\mbox{and}\quad j=1,\dots,k;
%\ee
\]

iv) if under the assumptions of (ii) and (iii)
the symmetries $Y_1, \dots,
Y_s$, where $s\leq r$, span an involutive distribution, i.e., $[Y_p,Y_q]=\sum\limits_{g=1}^s
c_{pq}^{g}(a_1,\dots,a_k, I_1,\dots,I_k,J_1,\dots,J_m) Y_g$ for all $p,q=1,\dots,s$
(the commutator is now given by (\ref{com_ode})!),
then the symmetries $\tilde Y_1, \dots,
\tilde Y_s$ also span an involutive distribution, $[\tilde Y_p, \tilde
Y_q]=\sum\limits_{g=1}^s \tilde c_{pq}^{g}\tilde Y_g$, for all
$p,q=1,\dots,s$, where $\tilde c_{pq}^{g}=c_{pq}^{g}(\tilde I_1,\dots,\tilde I_k, b_1,\dots,b_k,\tilde
J_1,\dots,\tilde J_m)$. If $c_{pq}^g\in\mathbb{K}$ are constants (in particular, they do not depend on $a_j$,
$j=1,\dots,k$), and thus
$Y_g$, $g=1,\dots,s$, form a Lie algebra, then $\tilde Y_g$, $g=1,\dots,s$, form an isomorphic Lie algebra.\looseness=-1
\end{cor}

{\bf Example 4.} Consider equation 6.45 from \cite{kamke}
\[
\frac{d^2 u}{dz^2}=c \left(\frac{du}{dz}\right)^2+a
\]
which admits an integral of the form
\[
I=((du/dz)^2+(a(1+2 c u))/(2 c^2))\exp(-2cu).
\]
Let $a_1\equiv a$ and $b_1\equiv b$. Then we have
\[
\tilde I=2c^2(b\exp(-2 c u)-(du/dz)^2)/(1+2cu),
\]
which is an integral for the transformed equation,
\[
d^2 u/dz^2=c (du/dz)^2(1-2 c^2/(1+2c u))+2b c^2\exp(-2 c
u)/(1+2cu).
\]

Just as for dynamical systems (\ref{ourds}), for ODEs we also
often can add parameters by hand through changes of variables, and
apply the generalized St\"ackel transform to the resulting
equations.

{\bf Example 5.} Consider equation 7.7 from \cite{kamke},
\[
\frac{d^3 u}{dz^3}=\frac{1}{u} \frac{d^2 u}{d z^2} \frac{du}{dz}-u^2 \frac{du}{dz},
\]
which admits an integral of the form
\[
\frac{1}{u} \frac{d^2 u}{d z^2}+\frac{u^2}{2},
\]
and rescale $u\rightarrow a u$.
This yields the equation
\[
\frac{d^3 u}{dz^3}=\frac{1}{u} \frac{d^2 u}{d z^2} \frac{du}{dz}- a u^2 \frac{du}{dz}
\]
with an integral %of the form
\[
I=\frac{1}{u} \frac{d^2 u}{d z^2}+\frac{a^2 u^2}{2}.
\]
Again let $a_1\equiv a$ and $b_1\equiv b$. Then we have
\[
\tilde I=\left(2\left(b-\frac{1}{u}\frac{d^2 u}{dz^2}\right)\right)^{1/2}\frac{1}{u},
\]
which is an integral for the transformed equation,
\[
\ds\frac{d^3 u}{dz^3}= \left(\frac{1}{u} \frac{d^2 u}{dz^2}- \left(2\left(b-\frac{1}{u}\frac{d^2 u}{dz^2}\right)\right)^{1/2} u\right) \frac{du}{dz}.
\]

\section{Conclusions and discussion}
In this paper we extend the
multiparameter generalized
St\"ackel transform, or the coupling constant
metamorphosis, to general dynamical systems (\ref{ourds}) and ODEs
and studied the properties of this extension. In particular, we present
sufficient conditions under which the transformed system
inherits the integrability properties of the original one: the
existence of Lax representation, (sufficiently many) integrals of
motion and symmetries, etc. In contrast with the Hamiltonian case
\cite{sb08}, for general dynamical systems (\ref{ourds}) we can avoid
introducing the reciprocal transformation for
(the solutions of) the equations of motion. As a result, the
relationship among the solutions of the original system
(\ref{ourds}) and the transformed system (\ref{trds}) is much
simpler than in the Hamiltonian case. Note that the same approach
was successfully applied to discrete dynamical systems, see
\cite{riq} for details.\looseness=-1

\looseness=-1

Our results naturally lead to a number of open problems related to
the generalized St\"ackel transform, of which we list below just a
few.

First of all, it would be very interesting to find out (both in
the Hamiltonian and the non-Hamiltonian case) when the transformed
dynamical system is {\em algebraically} \cite{v1,v2} completely
integrable provided so is the original system. On a related note,
the study of relationship among the differential Galois groups of
the variational equations (see e.g.\ \cite{vdps, az, mp}
and references therein for the relevant definitions)
for original and transformed systems would be of interest too.

Second, we have just barely scratched the surface by noticing in
Proposition~\ref{lax} that the transformed system inherits the
existence of a Lax representation from the original system, and
e.g.\ understanding what is the precise relationship among the
Darboux \cite{ms} and B\"acklund \cite{ks, kv} transformations for
the original and transformed system would certainly be worth the
while.

Third, it is highly desirable to study more systematically the
issue of when inserting the parameters `by hand' (cf.\ the above
Examples 2 and 5) and subsequent transforming of the resulting
systems leads to interesting new examples. We expect this
technique to yield a significant extension of the pool of exactly
solvable dynamical systems and ODEs. A good starting point here
could be e.g.\ to find the transformed counterparts for the ODEs
that linearize on differentiation \cite{fs}.

Finally, the simplest but perhaps also the most important (cf.\ e.g.\
\cite{sb08} for the Hamiltonian systems) special
case when integrals of motion are {\em linear} in the parameters
undoubtedly deserves to be explored in far more details.

We hope that the present paper will stimulate further research in
these and related areas.
\section*{Acknowledgments}

This research was supported in part
%by the Czech Grant Agency (GA\v{C}R) under grant No.\ 201/04/0538,
by the Ministry of Education, Youth and Sports of the Czech
Republic (M\v{S}MT \v{C}R) under grant MSM 4781305904 and
by the Czech Grant Agency (GA \v{C}R) under grant P201/11/0356.

It is my great pleasure to thank Prof.\ M. B\l aszak, Prof. A. Maciejewski,
and Prof.\ V.B.\ Matveev for stimulating
discussions, and Dr.\ R.O. Popovych for reading the manuscript of
the present paper and making many helpful comments. I also
thank the referees for useful suggestions.

\end{document}